\documentclass[aps,prb,twocolumn,groupaddress,letter]{revtex4}

\pdfoutput=1
\usepackage{hyperref}
\hypersetup{pdfauthor={KSD Beach, Fabien Alet, Matthieu Mambrini,  and Sylvain Capponi}, pdftitle={The SU(N) Heisenberg model on the square lattice: a continuous-N quantum Monte Carlo study}}

\usepackage{amsfonts,amssymb,amsmath}
\usepackage{color}
\usepackage{graphicx}
\usepackage{dcolumn}
\usepackage{bm}

\newcommand{\ket}[1]{|#1 \rangle}

\begin{document}

\title{The SU($N$) Heisenberg model on the square lattice: 
a continuous-$N$\\ quantum Monte Carlo study}

\author{K.\ S.\ D.\ Beach}
\email{kbeach@phys.ualberta.ca}
\affiliation{Department of Physics, University of Alberta, Edmonton, Alberta, Canada T6G 2G7}
\author{Fabien Alet}
\author{Matthieu Mambrini}
\author{Sylvain Capponi}
\affiliation{Universit\'e de Toulouse; UPS; Laboratoire de Physique Th\'{e}orique (IRSAMC); F-31062 Toulouse, France}
\affiliation{CNRS; LPT (IRSAMC); F-31062 Toulouse, France}

\date{September 22, 2009}

\begin{abstract}
A quantum phase transition is typically induced by tuning an external parameter that
appears as a coupling constant in the Hamiltonian. Another route is to vary the global symmetry 
of the system, generalizing, e.g., SU($2$) to SU($N$). In that case, however, the discrete 
nature of the control parameter prevents one from identifying and characterizing the
transition. We show how this limitation can be overcome for the SU($N$)
Heisenberg model with the help of a singlet projector algorithm that can
treat $N$ {\it continuously}. On the square lattice, we find a direct,
continuous phase transition between N\'{e}el-ordered and crystalline bond-ordered
phases at $N_{\text{c}}=4.57(5)$ with critical exponents
$z=1$ and $\beta/\nu = 0.81(3)$.
\end{abstract}


\maketitle

\section{Introduction}

The field of quantum magnetism encompasses a large variety of physical
phenomena that are of current experimental and theoretical interest.
These include competition between interactions (frustration), ordering in 
conventional or unconventional magnetic states, and the existence of
fractionalized excitations. 
In two dimensions, where some of the most unusual physics occurs,
there is a conspicuous absence of methods for  studying the behaviour of 
quantum magnets with high precision. On the analytical side,
neither the powerful methods devised for dimension $d=1$ (bosonization, 
conformal field theory) nor the mean-field methods exact in high $d$ are available. 
On the numerical side, simulations are difficult because of the
enormous size of the Hilbert space and, for stochastic methods, because of  the 
fatal ``sign problem.''

One way to relax these strong methodological constraints
is to decrease the role of quantum fluctuations. For instance, considering
the classical limit of magnets with large spin $S$ eases analytical studies. 
This limit, however, very often misses the
important competition between the instabilities existing only at
the quantum level. An alternative route---one that
preserves the quantum fluctuations---is to enlarge the
symmetry of the model, e.g., by extending the SU(2) spin symmetry
to SU($N$). This has proved very useful in the past, as the
$N\rightarrow \infty$ limit often allows for an exact analytical treatment. 
Methods to study $1/N$ corrections are also available, although they
cannot fully capture the exact details of what happens at finite $N$. 
The advantage of SU($N$) models over classical ones is that they naturally
allow for quantum states of matter (such as valence bond solids) by
construction, even if the off-diagonal elements of the Hamiltonian are
suppressed in the large-$N$ limit.

 Using this technique, and building on previous work,~\cite{Arovas88} Read and
Sachdev~\cite{ReadSachdev} have studied in detail the SU($N$) generalization
of the Heisenberg Hamiltonian on the square lattice. For sufficiently large $N$,
the system spontaneously breaks lattice translation symmetry
to form a valence bond crystal (VBC). For small $N$ [including the standard SU(2) model at $N=2$], 
the ground-state is antiferromagnetically ordered. 
The details of the phase diagram and of the VBC depend on the representation 
of the generators of the SU($N$) algebra considered:
for the case of square Young tableaux with $n$ columns, a
direct phase transition between the N\'eel and VBC states is predicted to
occur at the (mean-field) value $N/n\sim 5.26$. 

In a technical breakthrough, Kawashima and coworkers~\cite{Kawashima04,Harada03,Kawashima07} 
extended a quantum Monte Carlo (QMC) loop algorithm
designed for SU(2) models to the SU($N$) case (for all integer $N$ and for all
single-row Young tableaux). Studying the square lattice case with this exact numerical method, 
they found for $n=1$ that the $N=4$ model is N\'eel ordered, whereas
the $N=5$ model supports VBC order. This confirmed the analytical
large-$N$ predictions, but because of the discrete nature of the
algorithm, these studies could not rule out an intermediate phase between
$N=4$ and $5$. Even if this (possibly spin-liquid~\cite{Santoro}) phase does not exist, it
is impossible to obtain a precise value for the critical parameter $N_{\text{c}}$
separating the two phases and to ascertain the nature of the phase
transition at $N_{\text{c}}$. 

In this paper, we describe a quantum Monte Carlo algorithm, formulated in the total singlet
basis, that can treat the parameter $N$ continuously (in the manner of analytical, large-$N$ techniques). 
Applying this approach to the square-lattice SU($N$) Heisenberg model, we find that there is a direct transition occurring at
$N_{\text{c}}=4.57(5)$ between the N\'eel and VBC columnar phases. The transition is found to be second order,
with critical exponents $z=1$ and $\beta/\nu=0.81(3)$.
At the end of the paper we discuss  the implications of finding a second-order quantum phase
transition between states with incompatible symmetries and the possible connection to the deconfined
quantum criticality (DQC) scenario.~\cite{Senthil04}

\section{Model Hamiltonian}

\begin{figure}
\includegraphics{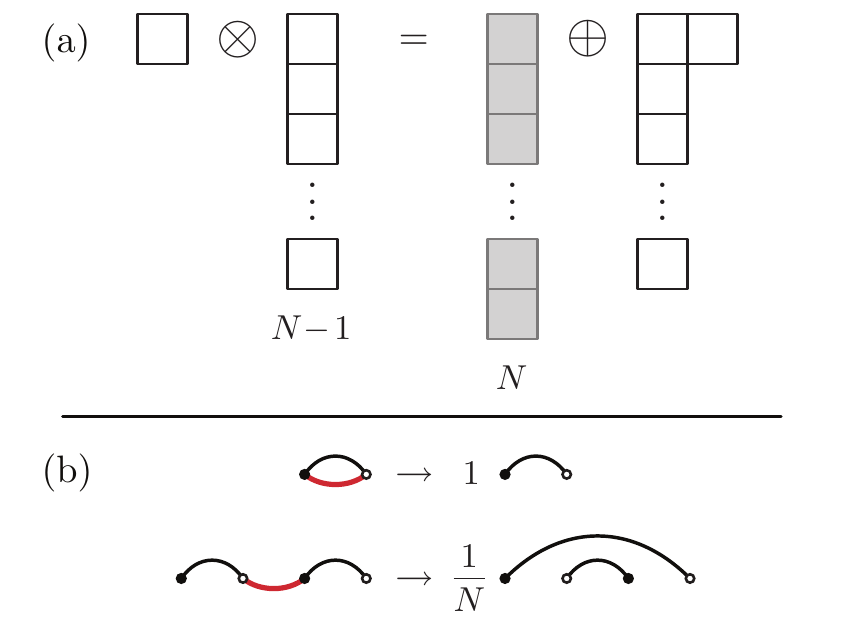}
\caption{(Color online) (a) An A-sublattice and a B-sublattice spin can pair to
form a singlet, which corresponds to a column of zero (modulo $N$) boxes.
 (b) Update rules for the action of $H_{ij}$ (red line) on VB
  singlet states (black bonds) .}
\label{fig:tableau}
\end{figure}

Our starting point is the SU($N$) generalization of the quantum Heisenberg model:
\begin{equation} \label{eq:H}
H=-J\sum_{\langle i,j\rangle}H_{ij}=\frac{J}{N}\sum_{\langle i,j\rangle}\sum_{\alpha,\beta=1}^{N}{\cal
  J}^\alpha_\beta(i){\cal J}^\beta_\alpha(j).
\end{equation}
Here, the exchange coupling $J=1$ sets the energy scale, and $\langle i,j\rangle$ denotes 
nearest-neighbor sites $i$ and $j$. The generators of the SU($N$) algebra, ${\cal J}^\alpha_\beta$,
satisfy the anticommutation relation
\begin{equation}
[J^\alpha_\beta(i),J^{\alpha'}_{\beta'}(j)]=\delta_{ij}(\delta_{\alpha\beta'}J^{\alpha'}_\beta\!(i)-\delta_{\alpha'\beta}J^{\alpha}_{\beta'}(i)).
\end{equation}
We consider the ``quark-antiquark'' model, taking the fundamental
representation of the generator on one sublattice (a single-box Young tableau)
 and its conjugate ($N-1$ boxes in one column) on the other. The fusion
 rule for the two representations is depicted graphically in Fig.~\ref{fig:tableau}(a).

If we describe the fundamental representation using a basis
$| \alpha \rangle$ with $\alpha = 1,2, \ldots, N$, then the conjugate 
representation has states $|\bar{\alpha}\rangle$ that are fully antisymmetrized 
tensor products of the form
\begin{equation}
|\bar{\alpha}\rangle = \frac{1}{\sqrt{(N-1)!}} \sum_{\alpha_2,\ldots,\alpha_N} \epsilon^{\alpha,\alpha_2,\ldots,\alpha_N}
|\alpha_2, \ldots, \alpha_N \rangle.
\end{equation}
An SU($N$) singlet formed between spins at sites $i$ and $j$ in opposite 
sublattices is the maximally entangled state
\begin{equation} \label{eq:SUNsinglet}
\begin{split}
|\phi\rangle_{ij} &= \frac{1}{\sqrt{N}} \sum_{\alpha = 1}^N |\alpha\rangle_i |\bar{\alpha}\rangle_j \\
&= \frac{1}{\sqrt{N!}} \sum_{\alpha_1,\ldots,\alpha_N} \epsilon^{\alpha_1,\alpha_2,\ldots,\alpha_N}
|\alpha\rangle_i |\alpha_2, \ldots, \alpha_N \rangle_j.
\end{split}
\end{equation}
Equation~\eqref{eq:H} can be understood as an operator
that performs a local singlet projection $H_{ij} = -\frac{1}{N}{\cal
  J}^\alpha_\beta(i){\cal J}^\beta_\alpha(j) =  |\phi\rangle_{ij}\langle\phi|_{ij}$
across all links of the square lattice.

The SU($N$) Heisenberg model can alternatively be written as an SU(2)
system with spin $S=(N-1)/2$ moments interacting via higher-order exchange processes.
An exact mapping connects the conventional SU(2) spin operators 
to the SU($N$) generators as follows:
\begin{align}
S^+ &= \sum_{\alpha=1}^{N-1} \sqrt{\alpha(N-\alpha)}\mathcal{J}_\alpha^{\alpha+1},\\
S^- &= \sum_{\alpha=1}^{N-1} \sqrt{\alpha(N-\alpha)}\mathcal{J}_{\alpha+1}^\alpha, \\ 
S^z &= \frac{1}{2}\sum_{\alpha=1}^N \bigl(N+1-2\alpha\bigr)\mathcal{J}_\alpha^\alpha.
\end{align}
The Hamiltonian can then be expressed in terms of
\begin{equation} \label{eq:Halt}
H_{ij} = \prod_{l=1}^{2S} \biggl[ 1 - 2\biggl(\frac{S(S+1)+{\bf S}_i\cdot{\bf S}_j}{l(l+1)}\biggr)\biggr].
\end{equation}
For instance, $H_{ij} =1/4-{\bf S}_i\cdot{\bf S}_j$ for $S=1/2$ ($N=2$); 
$H_{ij} =\frac{1}{3}[({\bf S}_i\cdot{\bf S}_j)^2-1]$ for $S=1$ ($N=3$); etc.
This was the starting point of previous finite-temperature QMC
investigations of this model,~\cite{Harada03,Kawashima07} where a path-integral technique was developed
in the $S_z$ basis of the spins.~\cite{Kawashima04} In this paper, we take a rather different route, using a $T=0$
algorithm formulated in the SU(2) total singlet basis of the spins $S$.

\section{Bond Basis}

Consider the subspace formed by \emph{bipartite valence bond} (VB) states~\cite{Beach06}
in which  two spins $S$ in opposite sublattices form a VB by coupling pairwise in a
singlet, formally given in the $S_z$ basis by 
\begin{equation}
|\phi\rangle_{ij}
=\frac{1}{\sqrt{2S+1}}\sum_{m=-S}^S (-1)^{m-S} {|m\rangle_i}\otimes{|-m\rangle_j}
\end{equation}
[cf.\ Eq.~\eqref{eq:SUNsinglet}].
For general $S$, this subspace does \emph{not} span the full  SU(2) singlet manifold, 
but only the subspace of states that are also SU($2S+1$) symmetric. 

For bipartite valence bonds, we can impose a VB orientation convention such that the 
overlap between any two states is positive. This basis is nonorthogonal, and the overlap 
between two VB states is 
\begin{equation}
\langle v_1 | v_2 \rangle=
(2S+1)^{N_l-N_v},
\label{eq:overlap}
\end{equation}
where $N_l$ is the number of {\it loops} formed by superimposing the two VB states
$\ket{v_1}$ and $\ket{v_2}$, and $N_v$ is the number of VBs in each state (or, equivalently, 
half the number of lattice sites). This is a simple generalization of the well-known overlap rule for $S=1/2$.

The Perron-Frobenius theorem tells us that on a bipartite, finite lattice
the SU($N$) symmetric Hamiltonian~\eqref{eq:H} admits a unique ground state. 
This state is an SU($N$) singlet, which can be expressed in the bipartite VB basis.

The operator $H_{ij}$ obeys the rules $H_{ij}|\phi\rangle_{ij} = |\phi\rangle_{ij}$ and 
$H_{ij}|\phi\rangle_{il}|\phi\rangle_{kj} = \frac{1}{N} |\phi\rangle_{ij}|\phi\rangle_{kl}$.
As a consequence, the action of $H_{ij}$ on VB states is extremely simple~\cite{Affleck90} and
consists of the bond rearrangements depicted in Fig.~\ref{fig:tableau}(b). 
We exploit this fact to simulate the SU($N$) model, noting that
the VB projector QMC (Ref.~\onlinecite{Sandvik05}) developed for $S=1/2$ works
with \emph{precisely such update rules}. For the sake of completeness, we describe this method 
(emphasizing the few details that differ) in Sec.~\ref{SEC:NumericalMethod}.

\begin{figure}
\includegraphics{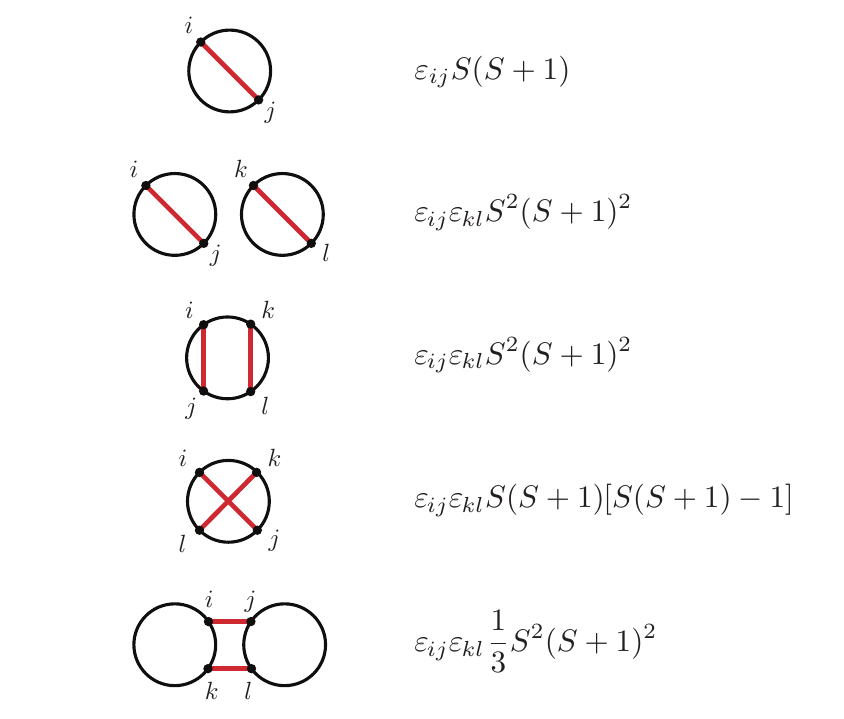}
\caption{(Color online) Loop diagrams contributing to two- and four-spin
correlators [${\bf S}_i \cdot {\bf S}_j $ and 
$({\bf S}_i \cdot {\bf S}_j) ({\bf S}_k \cdot {\bf S}_l) $, respectively] and their contributions. 
These overlap loops (schematically circular here) are obtained by superimposing the VB configurations
$\ket{v_1}$ and $\ket{v_2}$. The straight (red) lines link the corresponding sites of the correlation function. 
Only diagrams with nonvanishing contributions are displayed.
}
\label{fig:rules}
\end{figure}

What remains is to determine how to compute observables. It is
well-known in the $S=1/2$ case that most observables can be written in terms of estimators based
on the overlap loops.~\cite{Beach06} For instance, the spin correlator $\langle
v_1 | {\bf S}_i\cdot {\bf S}_j | v_2 \rangle/\langle v_1 | v_2
\rangle =(3/4) \epsilon_{ij} $ if spins $i$ and $j$ belong to the same loop,
$0$ otherwise. Here $\epsilon_{ij}=1$ if $i$ and $j$ are on the same
sublattice, $-1$ otherwise. We have generalized these rules for the spin
 $S$ case, and the resulting nonvanishing loop diagrams are given in
Fig.~\ref{fig:rules}, alongside the value of their contribution. 

We now make our key observation: since the update rules, the Monte Carlo weights, and the
estimators have analytical expressions in $N$ (or $S$, via $N=2S+1$),
simulations can be performed for {\it arbitrary, continuous values of $N$}. 
Formally, this can be understood as an analytic continuation from integer to real $N$. 
This is a great advantage over other QMC techniques,~\cite{Kawashima04} which are
restricted to half-integer and integer $S$. Large-$N$ analytical techniques~\cite{ReadSachdev} can also treat continuous values of $N$, but our numerical technique allows for an {\it
  exact} treatment of the Hamiltonian, in contrast to the mean-field
approximation inherent to the large-$N$ approach. 

\section{Numerical Method\label{SEC:NumericalMethod}}

\subsection{Quantum Monte Carlo algorithm}

We employ an efficient Monte Carlo algorithm that is an extension of
the VB projector scheme introduced by Sandvik.~\cite{Sandvik05}
The idea is to sample the ground state via the
power method by applying $(-H)^M$ (for fixed $M$ sufficiently large) to an arbitrary valence bond trial state:
\begin{equation}
| \psi_0 \rangle = \lim_{M\rightarrow \infty} (-H)^M | \varphi_T \rangle.
\end{equation}
For the purpose of sketching out the algorithm, 
it is convenient to rewrite Eq.~\eqref{eq:H} in a form that 
explicitly indexes the bonds that are acted upon:
\begin{equation} 
-H/J = \sum_{\langle ij\rangle} H_{ij} = \sum_b H_b.
\end{equation}
We make the equivalence $H_b = H_{i(b),j(b)}$, where
$b$ labels all the nearest-neighbour bonds on the square lattice. 
We now expand the powers of the Hamiltonian as
\begin{equation} \label{EQ:HM}
(-H)^M = \Bigl(\sum_b H_b\Bigr)^M=\sum_{\{b_1,b_2,...,b_M\}} \prod_{k=1}^{M} H_{b_k}.
\end{equation}
Each sequence $\{b_1,b_2,...,b_M\}$ corresponds to the process in which $| \varphi_T \rangle$ is acted on by 
local singlet projectors to give the propagated VB state $|\varphi_M\rangle$: 
\begin{equation}
|\varphi_M\rangle=H_{b_M}\ldots H_{b_2}H_{b_1} |\varphi_T\rangle.
\end{equation}

The sum over all possible sequences in Eq.~\eqref{EQ:HM} is evaluated stochastically.
For the SU($N$) model, the weight of each sequence is simply the product of the $1/N$ factors
(with $N=2$ for $S=1/2$) that appear as a result of the bond rearrangements [see Fig.~\ref{fig:tableau}(b)]
induced by the $H_b$'s. The most basic Monte Carlo move then consists of replacing a few $H_b$'s 
at random. Such changes are accepted or rejected depending on the ratio of the weight before and 
after the move.~\cite{Sandvik05}

To compute observables, two such sequences are generated with an additional factor in the sampling weight 
corresponding to the overlap of the two propagated states. For the SU($N$) model, this is computed 
according to Eq.~\eqref{eq:overlap}  The observables can then be measured using the loop 
estimators of Fig.~\ref{fig:rules} (see Sec.~\ref{SEC:Obervables}).

The spin gap $\Delta_{\text{s}}$, the energy difference between the first triplet excited state and the  singlet ground
state, can be computed using the same triplet propagation technique introduced for
the $S=1/2$ case.~\cite{Sandvik05,Beach06} Working with a single propagation sequence, we reinterpret the initial 
VB trial state as containing one triplet bond; the only difference in the propagation rules is that the triplet is annihilated
by $H_b$ when acted on directly [i.e., the coefficient $1$ in the first rule of Fig.~\ref{fig:tableau}(b) is replaced by $0$]. 
Looking at the statistics of the states that are not annihilated during the full propagation, one can easily compute the spin 
gap $\Delta_{\text{s}}$. We refer the reader to Refs.~\onlinecite{Beach06} and \onlinecite{Sandvik05} for more details.

To accelerate the convergence of the algorithm with respect to $M$, it is useful to sample a valence bond trial state 
that is a superposition of all valence bond configurations $v$ with amplitude $\varphi_T(v)$: 
\begin{equation}
| \varphi_T\rangle =\sum_v \varphi_T(v) \lvert v \rangle.
\end{equation}
There is no restriction on the trial state other than that $\varphi_T(v)$
be real and nonnegative and that ratios $\varphi_T(v_2)/\varphi_T(v_1)$ be easy
to compute for small changes in configuration $v_1 \to v_2$. In this work, we
make use of a simple RVB trial state~\cite{Liang88} in which the weight
$\varphi_T(v) = \prod_{[i,j] \in v} h_{ij}$ is a product of individual bond amplitudes
$h_{ij} = 1/r_{ij}^p$ that fall off algebraically as a function of the bond length.
The trial state is thus characterized by a single exponent $p$, which is a free
parameter in our simulations. To ensure that our results are fully converged
(i.e., that $M$ has been chosen sufficiently large) and do not 
show any residual dependence on the choice of $p$,
we have carried out the ground state projection starting
from four different RVB trial states (two magnetically ordered and two 
disordered) corresponding to $p=2.7$, 3.0, 3.5, and 5.0, and 
have checked that all observables converge to the same values.

We should point out that the updates are simple and sign-problem-free only
because $H_{ij}$, appearing in Eqs.~\eqref{eq:H} and \eqref{eq:Halt}, is
a pure, local-singlet projector. Our algorithm does not apply to
general spin-$S$ Hamiltonians. In any case, the bipartite VB states do not
form a basis for the singlet sector of general spin-$S$ models.

\subsection{Calculation of observables~\label{SEC:Obervables}}

The quantities of interest at the transition are the
staggered magnetization
\begin{equation}
{\bf M} = \frac{1}{L^2}\sum_{\bf r}
(-1)^{r_x+r_y}{\bf S}_{\bf r},
\end{equation}
its Binder cumulant
\begin{equation}
U=1-\frac{3 \langle {\bf M}^4 \rangle}{5\langle {\bf M}^2 \rangle^2},
\end{equation}
and the dimer order parameter ${\bf D} = (D_x, D_y)$, whose
two components
\begin{equation}
D_a = \frac{1}{L^2}\sum_{{\bf r}}(-1)^{r_a}{\bf S}_{\bf r}\cdot{\bf S}_{{\bf r}+\hat{\bf e}_a} \ \ (a = x,y)
\end{equation}
are directed along the square lattice vectors $\hat{\bf e}_x$ and $\hat{\bf e}_y$.

The measurements $\langle {\bf M}^2 \rangle$, $U$, and $\langle {\bf D}^2 \rangle$ are
carried out using the two- and four-point rules shown in Fig.~\ref{fig:rules}. 
For instance, using the same techniques as in Ref.~\onlinecite{Beach06}, we obtain  the following estimators:
\begin{equation}
\frac{\langle \varphi_M | {\bf M}^2 | \varphi_M' \rangle}{\langle \varphi_M | \varphi_M' \rangle}
=S(S+1)\sum_\alpha L^2_\alpha\end{equation}
and 
\begin{equation}
\begin{split}
\frac{\langle \varphi_M | {\bf M}^4 | \varphi_M' \rangle}{\langle \varphi_M | \varphi_M' \rangle} & = -\frac{1}{3}\bigl[S(S+1)+2S^2(S+1)^2\bigr] \sum_\alpha L^4_\alpha  \\
& \quad+  \frac{4}{3}S(S+1)\sum_\alpha L^2_\alpha\\
&\quad +\frac{5}{3}S^2(S+1)^2 \Bigl(\sum_\alpha L^2_\alpha\Bigr)^2,
\end{split}
\end{equation}
where the sum on $\alpha$ runs over all loops formed by the 
overlap $\langle \varphi_M | \varphi_M' \rangle$
of the two propagated states, and $L_\alpha$ is the size of a given loop.

\section{Results}

\begin{figure}
\includegraphics{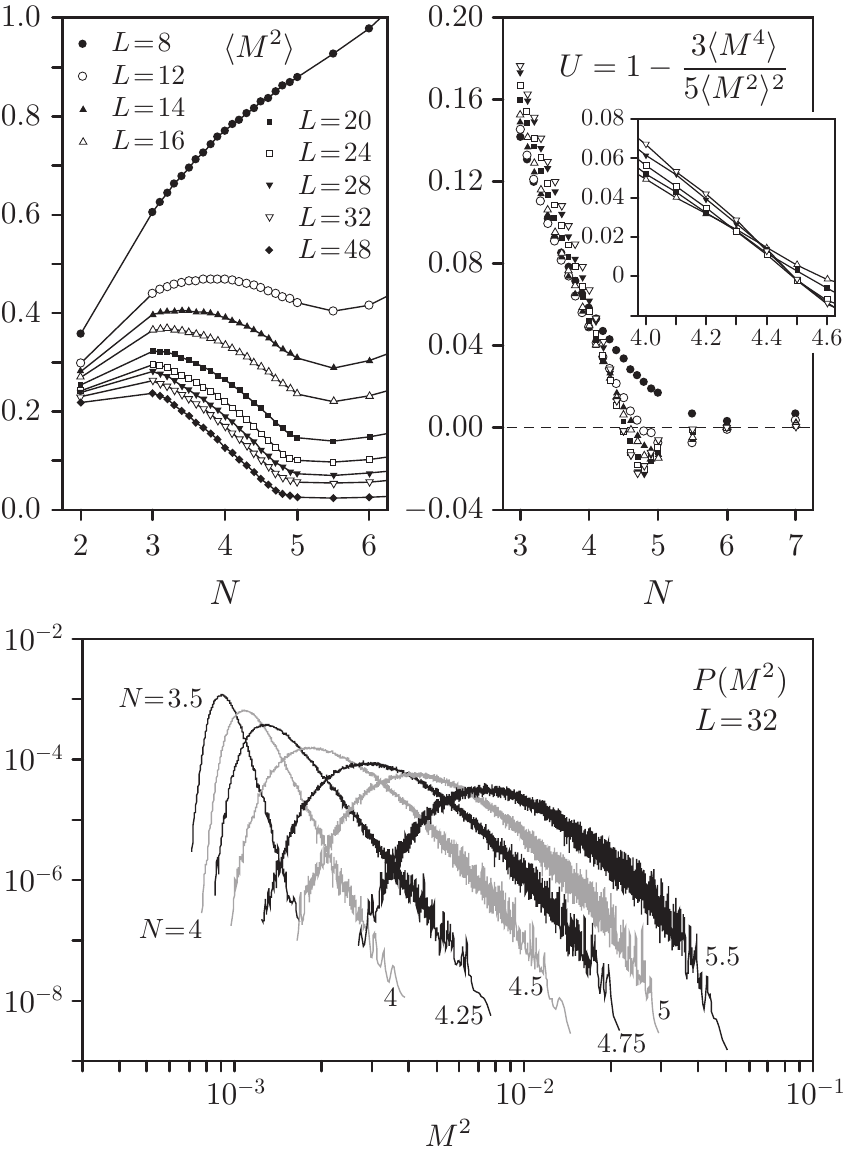}
\caption{(Top-left panel) Square of the staggered magnetization ${\bf M} = \frac{1}{L^2}\sum_{\bf r}
(-1)^{r_x+r_y}{\bf S}_{\bf r}$ as a function of $N$ for systems up to linear size $L=48$ (lines are guides to the eyes).
(Top-right panel) The Binder cumulant $U$ measures the kurtosis of the staggered magnetization
with respect to a purely Gaussian distribution. It vanishes in the limit $L\to \infty$ in the absence of antiferromagnetic order.
(Bottom panel) A histogram of the magnetic observable is shown for values of $N$ spanning the transition. 
The distribution has a single peak.
}
\label{fig:mag}
\end{figure}

We proceed by presenting our results for the square lattice SU($N$) model.
Figure~\ref{fig:mag} shows the square of the staggered magnetization
and its Binder cumulant. It is apparent that in the thermodynamic limit a dome of 
antiferromagnetic order survives for $1 < N < N_{\text{c}}$ with $N_{\text{c}}$
between 4 and 5. The Binder cumulant $U$ vanishes on the large-$N$ side
of the transition and appears to have a crossing point at $N \approx 4.3$,
which drifts rightward as $L$ increases. We do not have data on sufficiently
large systems to ascertain that the crossing is stable and can be unambiguously identified
as the critical point. As we will see, a finite-size scaling analysis provides a better
estimate of $N_{\text{c}}$.

\begin{figure}
\includegraphics{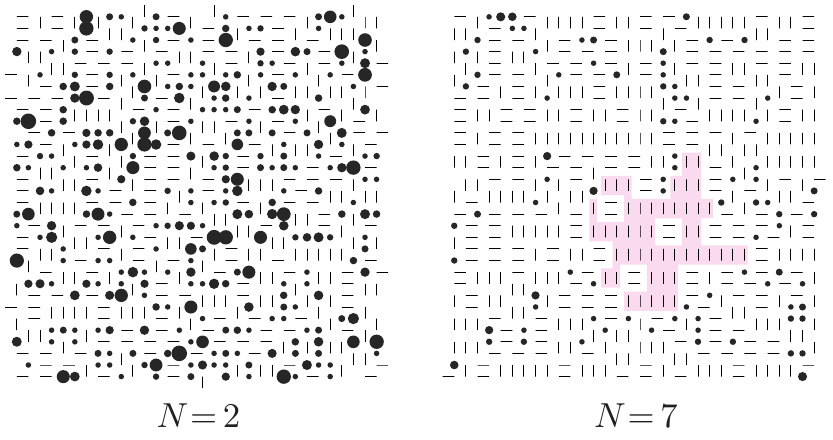}
\caption{(Color online) Snapshots of typical VB configurations for a system of size
$L=32$. Short, nearest-neighbour bonds are drawn with a line; long bonds are indicated
by circles (whose area is proportional to bond length) at their end points. 
For $N < N_{\text{c}}$, many long bonds stretch across the system.
For $N > N_{\text{c}}$, short bonds dominate. The shaded (pink) area marks a crystal domain with columnar bond order.
}
\label{fig:config}
\end{figure}

\begin{figure}
\includegraphics{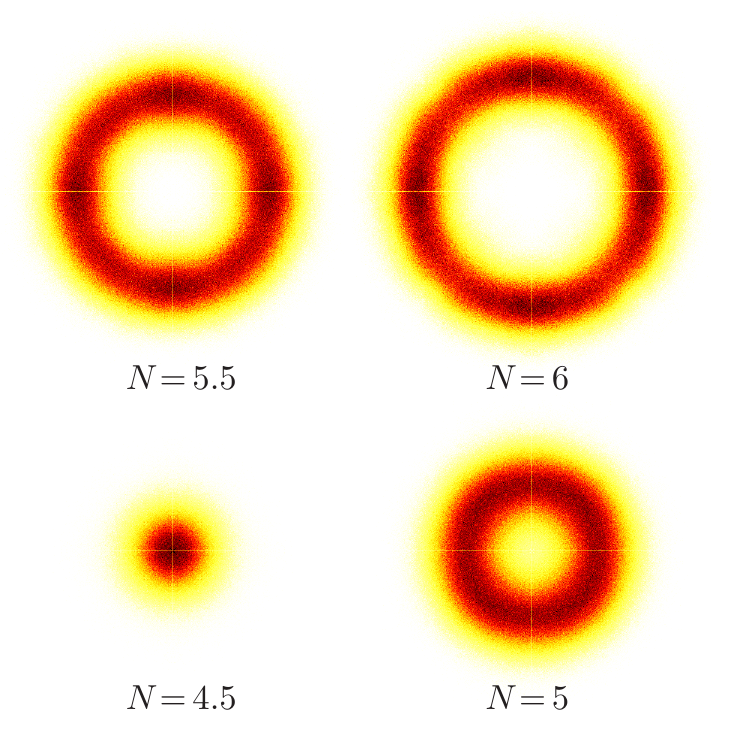}
\caption{(Color online) Density plots of histograms $P(D_x,D_y)$ of the $x$- and $y$-
components of the dimer order parameter ($L=32$). On the magnetic side of 
the transition (e.g., $N = 4.5 < N_{\text{c}}$), the distribution is characterized by a central peak. 
On the VBC side, the distribution is ring-like but develops additional weight along the
main axes as $N$ is increased.
}
\label{fig:ring}
\end{figure}

The destruction of the N\'{e}el order is driven by the gradual elimination
of singlet pairs that are correlated over long distances. In the large-$N$
phase, the singlets are predominantly short-ranged and form domains
of columnar ordering (see Fig.~\ref{fig:config}). 
On small lattices, these ordered domains are weak and they are almost equally
distributed among the four degenerate configurations. This can be seen in
the ring-like structure of the probability distribution $P(D_x,D_y)$, shown in Fig.~\ref{fig:ring}, 
that appears for $N > N_{\text{c}}$. Previous work~\cite{Harada03,Kawashima07} established the
existence of VBC order for $N\geq 5$, but could not determine whether it 
was of columnar [$\langle {\bf D}\rangle\!\sim\!(0,\pm D), (\pm D,0)$] or plaquette 
[$\langle {\bf D}\rangle\!\sim\!(\pm D,\pm D)$] symmetry. Our results suggest the former.
There does not appear to be a true U(1) degeneracy, as suggested in Ref.~\onlinecite{Kawashima07}.
Instead, it seems that there is an additional length scale $\xi_{\text{VBC}}$
much larger than the spin correlation length $\xi$. For $\xi < L < \xi_{\text{VBC}}$, there
is merely an effective U(1) degeneracy. 

\begin{figure}
\includegraphics{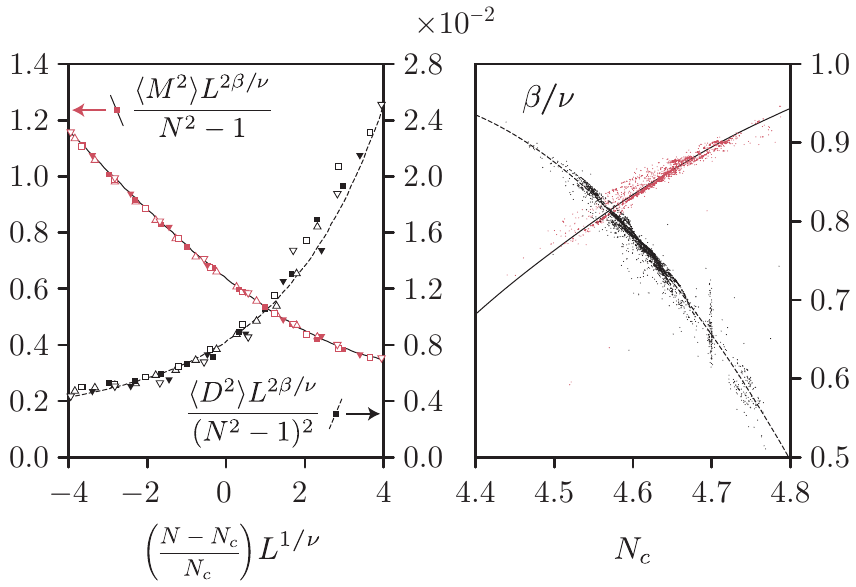}
\caption{(Color online)
Simultaneous data collapse of the N\'eel and dimer correlations
can be achieved with a single set of critical exponents. The left panel shows the QMC measurements
plotted in rescaled coordinates with the values $\nu=0.88$, $\beta=0.71$, and $N_{\text{c}} = 4.57(5)$.
Other values of $\beta/\nu \sim 0.8$ produce good data collapse.
The right panel shows bootstrapped results of $\beta/\nu$ versus $N_{\text{c}}$ when fits of 
$M^2$ and $D^2$ are performed independently. A single critical point corresponds to the crossing
$\beta/\nu=0.81(3)$ and $N_{\text{c}} = 4.57(5)$.
}
\label{fig:collapse}
\end{figure}

A finite-size scaling analysis of the magnetization and dimer order parameters suggests a
continuous transition defined by a single critical value $N_{\text{c}}$ and a single
set of critical exponents (see Fig.~\ref{fig:collapse}).
The data are not sufficiently sensitive to fix the exponent $\nu$ precisely,
 and the unusual behaviour of the Binder cumulant---its negative region and strong
subleading corrections---makes it unreliable for obtaining
an independent estimate of $\nu$. Reasonable fits seem to be achievable for a
range of values $0.75 \lesssim \nu \lesssim 1$. On the other hand, the ratio
$\beta/\nu$ is relatively stable. Repeating our fitting procedure for $M^2$ and $D^2$
independently with 3000 bootstrap samples, we conclude that both quantities vanish
simultaneously and continuously at $N_{\text{c}} = 4.57(5)$ with 
$\beta/\nu = 0.81(3)$. Note that the anomalous dimension $\eta = 2\beta/\nu - 1 = 0.63(4)$ is
at least an order of magnitude larger than what would be expected from either the
three-dimensional $O(3)$ or $\mathbb{Z}_4$ universality classes. 

\begin{figure}
\includegraphics{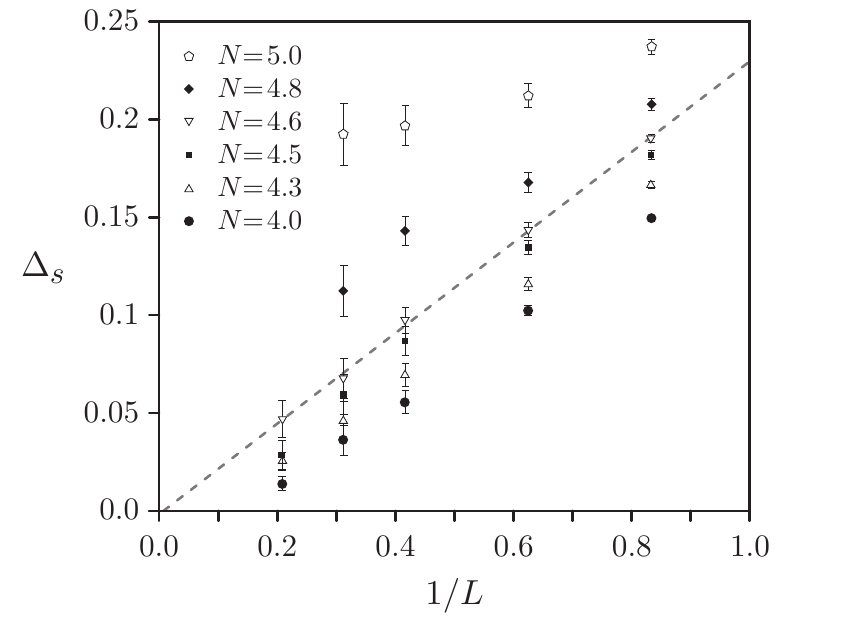}
\caption{The spin gap $\Delta_{\text{s}}$ versus the inverse linear size $1/L$ close to the critical point. 
The dashed line indicates a linear fit for $N=4.6$ that extrapolates very close to $0$ in the thermodynamic limit. 
}
\label{fig:gap}
\end{figure}

In Fig.~\ref{fig:gap}, we plot the spin gap as a function of $1/L$ in the region $4 \le N \le 5$. For 
$N$ large enough, the data converges to a nonzero value in the thermodynamic limit. 
On the other hand, for the smallest $N$, the gap clearly vanishes.
A linear extrapolation in $1/L$ (dashed lines in Fig.~\ref{fig:gap}) reveals that the gap closes at $N\approx 4.6$, in good
agreement with the value $N_c=4.57(5)$ derived from the order parameters. An unconstrained power-law 
fit yields a slope $1.02(10)$, confirming that the dynamical critical exponent of the phase transition is $z=1$. 

As is the case for any finite-lattice simulation, we cannot completely rule out an extremely weak 
first-order transition. Nonetheless, we have taken great care to examine the transition for signs 
of first-order character and have concluded that it is almost certainly continuous.  We have generated 
energy and order parameter histograms with extremely good statistics close to the transition and have
found no evidence of a bimodal character. See, e.g., the distribution of the staggered magnetization, 
shown in the bottom panel of Fig.~\ref{fig:mag}, which is single-peaked and evolves smoothly across 
the transition. 

One point of concern was that the Binder cumulant has a small region around the transition where 
it drops below zero. This is sometimes a signature of a first-order transition or of a distribution with a 
complicated multi-peaked structure. The histograms discussed above rule out the latter: the negative 
$U$ seems merely to correspond to a region in which the distribution of the magnetic order parameter 
is super-Gaussian (excess kurtosis). Another important observation relates to how the negative region 
evolves with system size. For first-order phase transitions, one typically observes very large negative 
values of the Binder cumulant that \emph{increase} in magnitude with the system size. This has been 
explained phenomenologically by Vollmayr et al.~\cite{Vollmayr93} For the transition observed here, 
the opposite is true: not only are the values only slightly negative, but a careful examination shows that 
the height of the dip begins to saturate starting around $L=28$. In addition to the depth of the negative 
region being bounded, its width also vanishes in the thermodynamic limit. 
Finally, it is also worth pointing out that the behaviour of the SU($N$) transition reported in this work 
bears little resemblance to the transition in models whose N\'{e}el-VBS transition is known to be 
first-order (e.g., Ref.~\onlinecite{Beach07}), which is marked by strong hysteresis effects
even at small lattice sizes.
These observations suggest 
to us that the behaviour is only a finite-size effect.

\section{Conclusion}

In conclusion, we have introduced an algorithm to simulate Heisenberg
SU($N$) models (for the representation with a single row and column on one sublattice and its conjugate on the other) on any bipartite lattice. It is formulated in the total singlet sector
and allows for efficient computation with arbitrary, continuous values
of $N$. For the square-lattice model, we find a second-order phase
transition between a N\'eel and VBC columnar state at
$N_{\text{c}}=4.57(5)$. 
Constructing a Ginzburg-Landau theory for this
phase transition is not simple from the symmetry point of view, as the
external parameter $N$ of the SU($N$) symmetry is tuned artificially
to unphysical values in our numerics.  Naively, the ingredients
seem similar to those encountered in DQC points~\cite{Senthil04,JQ}: a
continuous N\'eel-VBC transition, driven by an external parameter that
favors short VBs of the VBC over the long VBs needed for magnetic
ordering. Strictly speaking, the arguments of
Ref.~\onlinecite{Senthil04} do not apply here, since they rely heavily
on Berry phase effects specific to $S=1/2$. Hence, our results are not
directly related to those of Refs.~\onlinecite{Senthil04} and \onlinecite{JQ}.  The
large-$N$ techniques of Ref.~\onlinecite{ReadSachdev} do predict a
continuous transition from N\'eel order to disorder, but the ground state degeneracy can only be computed for integer $N$. 
We hope that our numerical results for the critical exponents encourage others to pursue
extended analytical calculations: the only available estimates of exponents~\cite{Halperin74} are for
representations with $n \gg 1$ and do not go beyond order $1/N$.

Finally, we note that the algorithm presented here can be applied with
minor modifications to the case of one-dimensional Heisenberg models with another
generalized symmetry, namely SU(2)$_k$. This open the door to the numerical
study of topological quantum liquids, as found in Ref.~\onlinecite{Feiguin07}.

During the completion of this work, a more efficient algorithm for the $N=2$ case 
was proposed by Sandvik and Evertz.~\cite{Sandvik08} This algorithm performs nonlocal 
moves by flipping spins around the loops in a mixed spin-VB representation of the projection.
It is straightforward to generalize this algorithm to the SU($N$) case for integer $N$, as has been done
in Ref.~\onlinecite{Lou09}. One of us~\cite{Kevinloop} has recently shown that an algorithm
for real $N$ can be constructed using a loop representation of  the SU($N$) model matrix elements 
similar to the one presented in Ref.~\onlinecite{Aizenman94}.
Early results obtained with this algorithm are in agreement with those presented in this work.

We thank K.\ Harada and N.\ Kawashima for fruitful exchanges. Some
calculations were performed using the ALPS libraries (Ref.~\onlinecite{ALPS}). We thank IDRIS and
CALMIP for allocation of CPU time. Support from the Procope (Egide), the French
ANR program under Grant No.\ ANR-08-JCJC-0056-01 (FA, MM, and SC) and 
from the Alexander von Humboldt foundation (KSDB) is acknowledged.


\begin{thebibliography}{99}

\bibitem{Arovas88} A.\ Auerbach and D.\ P.\ Arovas, Phys.\ Rev.\ Lett.\ {\bf 61},
  617 (1988); D.\ P.\ Arovas and A.\ Auerbach, Phys.\ Rev.\ B {\bf 38}, 316 (1988).

\bibitem{ReadSachdev} N.\ Read and S.\ Sachdev, Phys.\ Rev.\ Lett.\ {\bf 62},
  1694 (1989);  Nucl.\ Phys.\ {\bf B316}, 609 (1989); 
  Phys.\ Rev.\ B {\bf 42}, 4568 (1990).

\bibitem{Kawashima07} N.\ Kawashima and Y.\ Tanabe, Phys.\ Rev.\ Lett.\ {\bf 98},
  057202 (2007).

\bibitem{Harada03} K.\ Harada, N.\ Kawashima and M.\ Troyer,
  Phys.\ Rev.\ Lett.\ {\bf 90}, 117203 (2003).

\bibitem{Kawashima04} N.\ Kawashima and K.\ Harada, J.\ Phys.\ Soc.\ Jap.\ {\bf
  73}, 1379 (2004). 

\bibitem{Santoro} G.\ Santoro, S.\ Sorella, L.\ Guidoni, A.\ Parola, and E.\ Tosatti, 
Phys.\ Rev.\ Lett.\ {\bf 83}, 3065 (1999).

\bibitem{Senthil04} T.\ Senthil, L.\ Balents, S.\ Sachdev, A.\ Vishwanath, and M.\ P.\ A.\ Fisher, 
Science {\bf 303}, 1490 (2004); Phys.\ Rev.\ B {\bf 70}, 144407 (2004); J.\ Phys.\ Soc.\ Jpn.\ Suppl.\ {\bf 74}, 1 (2005). 

\bibitem{Beach06} K.\ S.\ D.\ Beach and A.\ W.\ Sandvik, Nucl.\ Phys.\ B {\bf 750}, 142 (2006).

\bibitem{Affleck90} I.\ Affleck, J.\ Phys.: Condens.\ Matter {\bf 2}, 405 (1990).

\bibitem{Sandvik05} A.\ W.\ Sandvik, Phys.\ Rev.\ Lett.\ {\bf 95}, 207203 (2005).

\bibitem{Liang88} S.\ Liang, B.\ Dou\c{c}ot, and P.\ W.\ Anderson, Phys.\ Rev.\ Lett.\ {\bf 61}, 365 (1988).

\bibitem{Vollmayr93} K.\ Vollmayr, J.\ D.\ Reger, M.\ Scheucher, and K.\ Binder, Z.\ Phys.\ B {\bf 91}, 113 (1993).

\bibitem{Beach07} K.\ S.\ D.\ Beach and A.\ W.\ Sandvik, Phys.\ Rev.\ Lett.\ {\bf 99}, 047202 (2007).

\bibitem{JQ}  O.\ I.\ Motrunich and A.\ Vishwanath, Phys.\ Rev.\ B {\bf 70}, 075104 (2004); A.W. Sandvik,  
Phys.\ Rev.\ Lett.\ {\bf 98}, 227202 (2007); R.G. Melko and R.\ K.\ Kaul, Phys.\ Rev.\ Lett.\ {\bf 100}, 017203 (2008); 
F.-J.\ Jiang, M.\ Nyfeler, S.\ Chandrasekharan, and U.-J.\ Wiese,  J.\ Stat.\ Mech.\ P02009 (2008)

\bibitem{Halperin74} B.\ I.\ Halperin, T.\ C.\ Lubensky, and S.-K.\ Ma, Phys.\ Rev.\ Lett.\ {\bf 32}, 292 (1974);
V.\ Y.\ Irkhin, A.\ A.\ Katanin, and M.\ I.\ Katsnelson, Phys.\ Rev.\ B {\bf 54}, 11953 (1996).

\bibitem{Feiguin07} A.\ Feiguin, S.\ Trebst, A.\ W.\ W.\ Ludwig, M.\ Troyer, A.\ Kitaev, Z.\ Wang,
and M.\ H.\ Freedman, Phys.\ Rev.\ Lett.\ {\bf 98}, 160409 (2007).

\bibitem{Sandvik08} A.\ W.\ Sandvik and H.\ G.\ Evertz, arXiv:0807.0682.

\bibitem{Lou09} J.\ Lou, A.\ W.\ Sandvik, N.\ Kawashima, arXiv:0908.0740.

\bibitem{Kevinloop} K.\ S.\ D.\ Beach, (unpublished).

\bibitem{Aizenman94} M.\ Aizenman and B.\ Nachtergaele, Comm.\ Math.\ Phys.\ {\bf 164}, 17 (1994).

\bibitem{ALPS} F.\ Albuquerque, F.\ Alet, P.\ Corboz, P.\ Dayal, A.\ Feiguin, S.\ Fuchs, L.\
Gamper, E.\ Gull, S.\ GŸrtler, A.\ Honecker, R.\ Igarashi, M.\ K\"{o}rner, A.\ Kozhevnikov,
A.\ L\"{a}uchli, S.\ R.\ Manmana, M.\ Matsumoto, I.\ P.\ McCulloch, F.\ Michel, R.\ M.\
Noack, G.\ Paw\l owski, L.\ Pollet, T.\ Pruschke, U.\ Schollw\:{o}ck, S.\ Todo, S.\ Trebst, and
M.\ Troyer, J.\ Magn.\ Magn.\ Mater.\ {\bf 310},
  1187 (2007); M.\ Troyer, B.\ Ammon and E.\ Heeb, Lect.\ Notes Comput.\ Sci., {\bf
  1505}, 191 (1998); see {\tt http://alps.comp-phys.org}.

\end{thebibliography}
\end{document}